\pdfoutput=1

\documentclass{article}

\usepackage{microtype}
\usepackage{graphicx}
\usepackage{booktabs}
\usepackage{array}
\usepackage{tabularx}
\usepackage{hyperref}
\usepackage{pdfpages}

\usepackage[accepted]{icml2026}

\makeatletter
\renewcommand{\Notice@String}{Accepted at ICML 2026 AI for Science Workshop.}
\makeatother

\usepackage{amsmath}
\usepackage{amssymb}

\icmltitlerunning{Grounded autonomous scrutiny at scale}

\begin{document}

\twocolumn[
\icmltitle{Grounded autonomous scrutiny at scale: \\
emergent critique from reproduction of published computational physics papers}

\icmlsetsymbol{equal}{*}

\begin{icmlauthorlist}
\icmlauthor{Haonan Huang}{pu}
\end{icmlauthorlist}

\icmlaffiliation{pu}{Department of Physics, Princeton University}

\icmlcorrespondingauthor{Haonan Huang}{hnhuang@princeton.edu}

\icmlkeywords{autonomous scientific agents, LLM agents, computational physics, density functional theory, scientific reproducibility, peer review}

\vskip 0.3in
]

\printAffiliationsAndNotice{}

\begin{abstract}
Autonomous LLM agents now produce complete research artifacts in machine-learning sandboxes, but real computational physics is harder: experiments are first-principles calculations against re-runnable physical ground truth, and meaningful new work almost always builds on a key existing paper. We ask whether such an agent can perform \emph{grounded scrutiny} of published computational physics --- reading a paper, reproducing it from scratch, and surfacing methodological concerns from execution. We deploy a single Claude Opus~4.6 configuration at two complementary scopes. At scale, across 111 open-access Quantum ESPRESSO papers, an autonomous agent runs the read--plan--compute--compare loop and, although never asked to critique, raises substantive methodological concerns on ${\sim}42\%$ of papers; \textbf{85 of 88 of these critiques (96.6\%) surface only after the agent has actually run a calculation, with a reading-only ceiling of 1.8\%}. Critique emerges from reproduction, not from reading. In depth, on one \emph{Nature Communications} paper on multiscale device simulation of a 2D-material MOSFET, a fresh agent inheriting a verified reproduction pipeline autonomously produces a 14-concern physics inventory and a complete, submission-form six-page Comment that revises the paper's $L_\text{G} = 5$~nm headline. Two of its $L_\text{G} = 5$~nm headline-challenging attacks --- a source-degeneration contact-resistance bound and a Sb-doping degradation ratio --- are absent from the published 21-reviewer peer review.
\end{abstract}

\section{Introduction}
\label{sec:intro}

Recent work has shown that an autonomous LLM agent can navigate the entire research life cycle of a machine learning project --- ideating, coding, running the training experiment, analyzing data, and writing the manuscript~\cite{lu2026}. The natural next question is whether the same kind of system can take on real-world physical science, where experiments are not training loops but first-principles calculations of real physical systems. \textbf{A fundamental difference between sandboxed automated research and real physical science is that the latter is computationally and intellectually demanding --- physics reasoning that cannot be reduced to interpolation, multi-scale calculations on decades-mature scientific software, and verifiability against re-runnable physical ground truth --- so meaningful new work almost always begins from a key existing paper.} A researcher reads the relevant literature, reproduces the central calculations in their own setup, critically evaluates what those calculations show, conceives what is missing or wrong, runs follow-up calculations to test the new idea, and possibly writes a publishable document. We refer to this stance --- audit by execution against the same physics the literature describes --- as \emph{grounded scrutiny}.

In this work we ask whether an autonomous LLM agent can perform grounded scrutiny on its own, in computational physics --- the natural testbed because numerical claims can be independently re-computed against the same physics, and community effort has established reproducibility standards for the simulations themselves~\cite{lejaeghere2016,bosoni2024}. We focus on density functional theory (DFT) and the Quantum ESPRESSO ecosystem~\cite{giannozzi2009,giannozzi2017}, and deploy a single Claude Opus~4.6 configuration at two complementary scopes: \textbf{at scale}, where a fresh agent reproduces an arbitrary published paper end-to-end; and \textbf{in depth}, where a single agent is unleashed on one carefully chosen paper to push reproduction-driven scrutiny as far as it will go.

At scale, across 111 open-access Quantum ESPRESSO papers, the agent autonomously runs the read--plan--compute--compare loop and reproduces roughly three-quarters of in-scope claims within 5\% of the published value. Unprompted, it raises substantive methodological concerns on ${\sim}42\%$ of papers --- \textbf{96.6\% of which require execution to surface, and only 1.8\% are catchable by reading alone}. \emph{Critique emerges from reproduction, not from reading}: critical scientific scrutiny is execution-bound, not a property of prior knowledge.

In depth, for one \emph{Nature Communications} paper on multiscale device simulation of a 2D-material MOSFET~\cite{pizzi2016}, a single agent goes well beyond reproduction. Given a verified multi-code reproduction pipeline, in one unsupervised session it inventories its physics concerns about the paper, runs three classes of new calculation the original work did not perform, and produces a complete, submission-form six-page Comment --- composed, figured, typeset, and PDF-iterated entirely by itself --- whose two main findings revise the paper's $L_\text{G} = 5$~nm headline conclusion. Neither finding appears in the published peer review, a complementarity we examine in \S\ref{sec:review}. \textbf{Unlike prior work in which an LLM reads a paper and flags errors by reading alone~\cite{son2025}, every numerical claim in our Comment is backed by a first-principles calculation the agent itself runs against the same physics the original paper used.} Scrutiny here is computational, not textual.

The capabilities exercised --- multi-scale physics reasoning, autonomous execution against re-runnable ground truth, and synthesis into a complete, submission-form artifact --- are precisely those a full real-world research loop demands; grounded scrutiny is the natural first stage of that loop and the central object of this paper.

\section{Grounded scrutiny at scale}
\label{sec:scale}

\subsection{Setup}

\begin{figure*}[!tb]
\centering
\includegraphics[width=\textwidth]{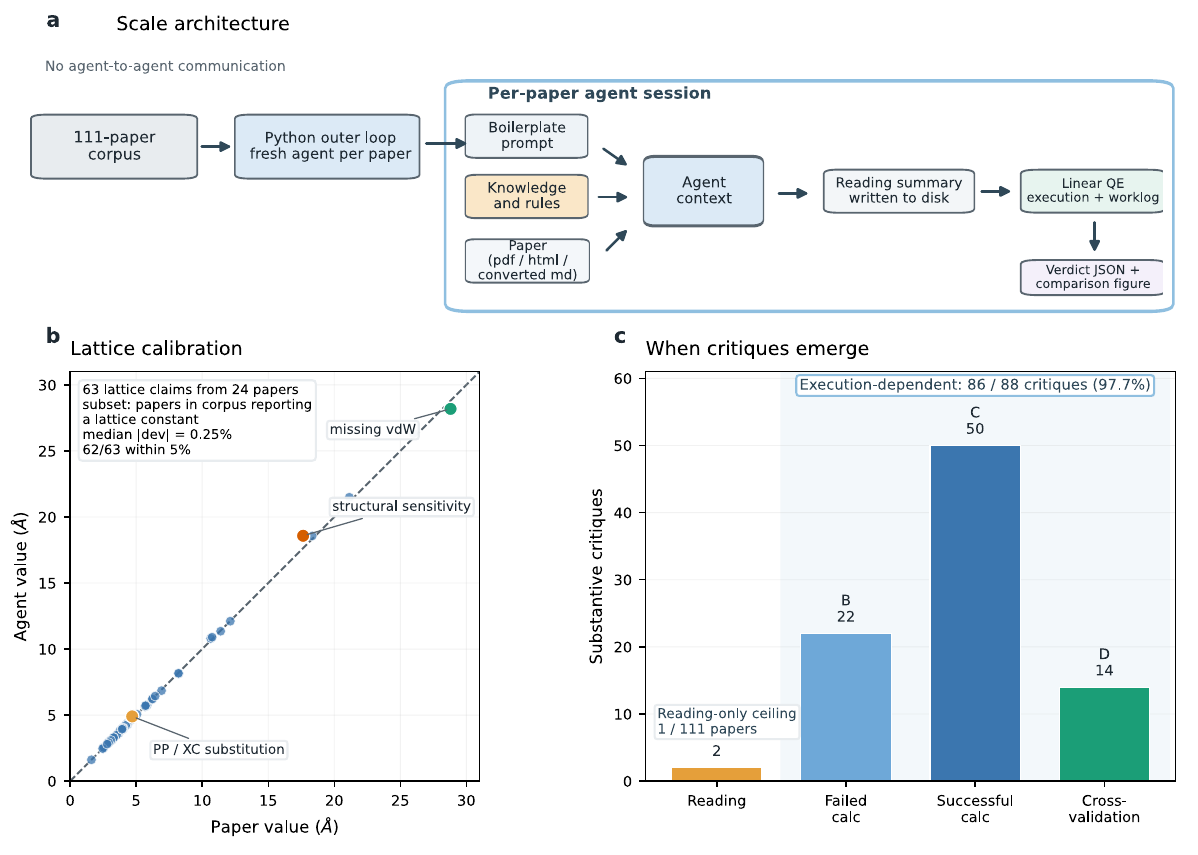}
\caption{\textbf{Grounded scrutiny at scale: architecture, calibration, and execution-dependence.}
\textbf{(a)}~Two-level pipeline: a Python outer loop iterates over the 111-paper corpus, handing each paper to a fresh Claude Opus~4.6 agent running inside the Claude Code CLI; there is no agent-to-agent communication. Within each per-paper session, three fixed inputs --- a boilerplate task prompt, a required-reading envelope of knowledge and house rules, and the paper itself --- feed the agent's working context, from which the agent writes a reading summary to disk, executes QE serially alongside a worklog, and emits a verdict JSON and comparison figures.
\textbf{(b)}~Lattice-constant calibration on the 24-paper subset reporting a lattice constant (63 claims), agent vs paper on a 45\textdegree{} line; median $|$dev$|$ = 0.25\%, 62/63 within 5\%. Outliers are labelled by dominant systematic: PP/XC substitution, missing vdW, structural sensitivity.
\textbf{(c)}~Phase classification of the 88 substantive critiques: A\,=\,3 (during reading), B\,=\,22 (after a failed calculation), C\,=\,49 (after a successful calculation compared to the paper), D\,=\,14 (after cross-validation). Execution-dependent: 85/88 = 96.6\%. Reading-only ceiling: 2/111 papers = 1.8\%. Caption counts are the corrected ledger (one row reclassified per the no-calculation clause; Appendix~\ref{app:methods} M6); the panel is drawn from the pre-correction ledger (A = 2, C = 50, 86/88, 1/111) and is superseded by these counts.}
\label{fig:fig1}
\end{figure*}

The corpus is \textbf{111 unique Quantum ESPRESSO papers} in which QE is the primary computational tool, filtered from all open-access QE literature~\cite{giannozzi2009,giannozzi2017} published 2010--2024 across 12 journal families (Appendix~\ref{app:methods} M1); restricting to open-access literature and open-source software lets us release every input, output, and trace. Across both the scale and depth regimes, the harness is the same: the \textbf{Claude Code CLI} as agentic orchestrator with \textbf{Claude Opus~4.6} as the underlying model, shelling out to bash for QE, Wannier90~\cite{pizzi2020}, and any analysis the agent writes itself in Python. There is no central tool layer --- no MCP server, no library wrapper --- by deliberate choice, to keep the harness honest about what the model does unaided. This differs from contemporary DFT-agent systems~\cite{wang2025dreams,kumar2026,zou2025} that serve as execution layers for human-specified tasks; none reads a published paper. The scale-mode pipeline (Fig.~\ref{fig:fig1}a) wraps this harness in a Python outer loop that iterates over the corpus, handing each paper to a fresh agent under a boilerplate prompt that converged in development and is then held fixed for production (Appendix~\ref{app:methods} M2, Appendix~\ref{app:prompts}). The agent receives the paper plus a small knowledge envelope covering core QE and the Wannier90 ecosystem.

Each agent is told to do four things in order: load the \textbf{full paper into context} --- information is interleaved across a paper in ways that require coherent understanding rather than discrete retrieval; write a structured reading summary with its plan and targets; execute calculations serially while updating a worklog; and emit a structured verdict. A wall-clock soft cap of \textbf{2--4~h} per paper is set as a flexibility budget rather than a hard timeout (Appendix~\ref{app:methods} M2); agents declare their own scope.

\subsection{Aggregate reproduction quality}
\label{sec:aggregate}

Within the agent's capability envelope, reproduction is uniformly strong. Papers are graded on a four-tier scale (T4 exact, T3 qualitatively correct, T2 partial, T1 failure); \emph{in-scope} is what fits the time budget, and agents declare scope in the verdict (Appendix~\ref{app:methods} M4). Across \textbf{571 deduplicated quantitative claims}, the agent matches \textbf{75.8\% within 5\%} and \textbf{83.2\% within 10\%} of the published value, with a \textbf{median deviation of 0.9\%}. At the paper level, 6 of 111 reach T4 and \textbf{90.9\% of the 110 scorable papers} reach T3 or T4. Figure~\ref{fig:fig1}b shows lattice constants across a 24-paper subset (63 claims): median $|$dev$|$ 0.25\%, 62/63 within 5\% --- the cleanest single-quantity calibration.

Two small knowledge files were nevertheless operationally important. Early runs frequently produced false refusals --- agents prematurely concluding ``QE cannot do this'' and skipping calculations they should not have. We tested this with a controlled ablation on 15 open-access papers under otherwise identical conditions, adding two compact text files (QE command idioms; pseudopotential-selection heuristics~\cite{prandini2018}) and a rule requiring the agent to consult them before declaring a capability absent. The false-refusal class was eliminated and attempted-workflow breadth expanded materially: phonon workflows attempted \textbf{rose from 3/15 to 5/15 papers (20\% $\to$ 33\%)} on the same set. The agent did not become more capable; it stopped declining capabilities it already had. \textbf{Two small text files unlock cognitive scope the model already has: the binding constraint for this failure class is the knowledge harness, not model capability.}

\subsection{Agent behaviour during a session}

Full trace analysis (Appendix~\ref{app:traces}) reveals a uniform \textbf{load-then-execute-serially} pattern: agents ingest paper and knowledge files first, write the reading summary, then interleave QE-input creation with execution, closing with the verdict. Three observations matter for \S\ref{sec:emergent}. Once the paper is in context, agents essentially never re-read it (nearly all sessions show zero late reads), treating the initial load and reading summary as sufficient working state. Under context compaction (16\% of sessions), agents rely on their on-disk QE inputs, outputs, and worklog rather than re-reading --- \textbf{the filesystem becomes external memory}, an emergent strategy we did not prompt for. And agents are \textbf{open-loop on visual output}: in zero of 61 inspected sessions does an agent open a figure it just wrote, notice a problem, and regenerate. We return to this in \S\ref{sec:discussion}.

\subsection{Emergent scrutiny and the execution requirement}
\label{sec:emergent}

Nothing in the production prompt asks the agent to critique the paper; the instruction is to reproduce, and deliverables are reproduction artifacts. With that caveat, the spontaneous critique rate is striking: \textbf{${\sim}42\%$ of papers (47/111) contain at least one substantive methodological concern} raised by the agent unprompted, across eight substantive categories (a ninth tracked category, agent-side convergence issues, is excluded; Appendix~\ref{app:vignettes}). A \emph{substantive methodological concern} is a paper-side concern under the Appendix~\ref{app:vignettes} taxonomy; agent-side convergence issues do not count. Denominators, defined once: 111 = the full deduplicated open-access corpus; 90 = the deduplicated union of unique papers across converged-harness batches B3--B5 (47/90 = 52.2\%); 50 = the controlled single-machine B5 run (30/50 = 60.0\%; Appendix~\ref{app:vignettes}). The total number of substantive critiques across the corpus --- a paper can contribute more than one --- is \textbf{88}.

The more interesting question is \emph{when} these critiques surface. Figure~\ref{fig:fig1}c classifies each post hoc by phase: during reading~(A), after a failed calculation~(B), after a successful calculation compared to the paper~(C), or after cross-validation~(D). Counts: \textbf{A\,=\,3, B\,=\,22, C\,=\,49, D\,=\,14} (corrected ledger; Appendix~\ref{app:methods} M6).

\textbf{Eighty-five of eighty-eight critiques (96.6\%) emerged only after the agent had actually run a calculation} (coding-sensitivity analysis: Appendix~\ref{app:methods} M6). The three Phase-A critiques are one file-corruption artifact and two genuine reading catches: a phosgene adsorption energy reported at 142~Ry, three-to-five orders beyond physical, and a reproducibility-omission catch --- a paper whose structure coordinates are omitted from both text and SI, the structure being available only behind a subscription --- which is not a physics catch like the phosgene case. The reading-only ceiling is \textbf{2 in 111 papers (1.8\%)}. Phase~C is the modal discovery point --- the agent runs, compares, notices --- and Phase~B is the error-recovery path where diagnosing a failed run reveals a paper-side concern. The 96.6\% execution requirement complements passive-reading verification baselines (${\sim}21\%$ recall on SPOT's 91 retraction/errata-level errors~\cite{son2025}) and code-provided reproduction baselines in the 19--21\% range (CORE-Bench~\cite{siegel2024}, PaperBench~\cite{starace2025}, ReplicationBench~\cite{ye2025}). To our knowledge this is the \textbf{first quantitative evidence that critical scientific scrutiny is execution-bound in an autonomous-agent corpus of this size}. On a 12-paper two-machine cross-check with no shared state, 7/12 produced overlapping critique category sets (5 substantive, 2 convergence-only) --- qualitatively consistent scrutiny patterns across independent runs.

\subsection{Inside the envelope}

Within the envelope, agents execute multi-code workflows autonomously across DFT+U, anomalous Hall conductivity, spin--orbit coupling, LDA+U correlated magnetism, \texttt{epsilon.x} optics, and DFPT dielectrics --- a six-panel agent-vs-paper diversity gallery is in Figure~\ref{fig:figA1} (Appendix~\ref{app:gallery}). Representative autonomous depth within one session includes full SCF $\to$ NSCF $\to$ Wannierization $\to$ MLWF hopping pipelines~\cite{marzari2012} on infinite-layer nickelates (nearest-neighbour hopping to 1.1\%, bandwidths to 0.3\%), and 44-MLWF Wannierizations with Berry-curvature integration~\cite{wang2006} recovering Mn$_3$Al $\sigma_{xy}$ within 4\%. \textbf{Scale mode is not running \texttt{pw.x} and reading off a gap} --- it autonomously executes the QE ecosystem as the paper demands.

Spontaneous discoveries have a concrete texture (four per-phase vignettes in Appendix~\ref{app:vignettes}): on a WS$_2$ monolayer paper (Phase~C), the agent's fully-relativistic calculation gave spin--orbit splitting 429~meV against the paper's 571~meV, with MoS$_2$ reproducing the paper within 2\% as an internal control and experiment (400--410~meV, supplied post hoc) agreeing with the agent. The envelope has limits scale alone cannot cross --- the 2--4~h budget and the QE + Wannier90 knowledge scope, both prices of a uniform pipeline across a hundred-plus papers. Can an agent go deep on a single paper, all the way to a physically meaningful verdict and on to a publication-shaped scientific artifact?

\section{From scale to depth: a test case on the boundary}

One paper in the corpus sits exactly on that boundary: Pizzi et al., arsenene and antimonene double-gate MOSFETs (\emph{Nat.\ Commun.}, 2016)~\cite{pizzi2016}, a first-principles device study arguing that 2D As and Sb form ultra-scaled sub-10~nm MOSFETs whose performance meets the ITRS industry roadmap. The paper combines four codes --- QE~\cite{giannozzi2009,giannozzi2017}, Wannier90~\cite{pizzi2020}, NanoTCAD~ViDES~\cite{marian2023,fiori2005}, and custom post-processing --- across DFT structure and band structure, ballistic NEGF transport, phonon scattering, and device figures of merit. Its central claim, on which its scientific interest rests, is the sub-10~nm ITRS-compliance of these devices. In scale mode the agent reproduces the QE-accessible claims at strong fidelity and stops at Wannier90, because NanoTCAD~ViDES sits outside the knowledge envelope. Pizzi~2016 is close enough to the envelope that the agent gets most of the way autonomously, and far enough beyond it that completing the chain exercises depth-mode grounded scrutiny end-to-end.

\section{Reproduce--Review--Reflect: depth-mode grounded scrutiny}
\label{sec:depth}

\begin{figure*}[!tb]
\centering
\includegraphics[width=0.82\textwidth]{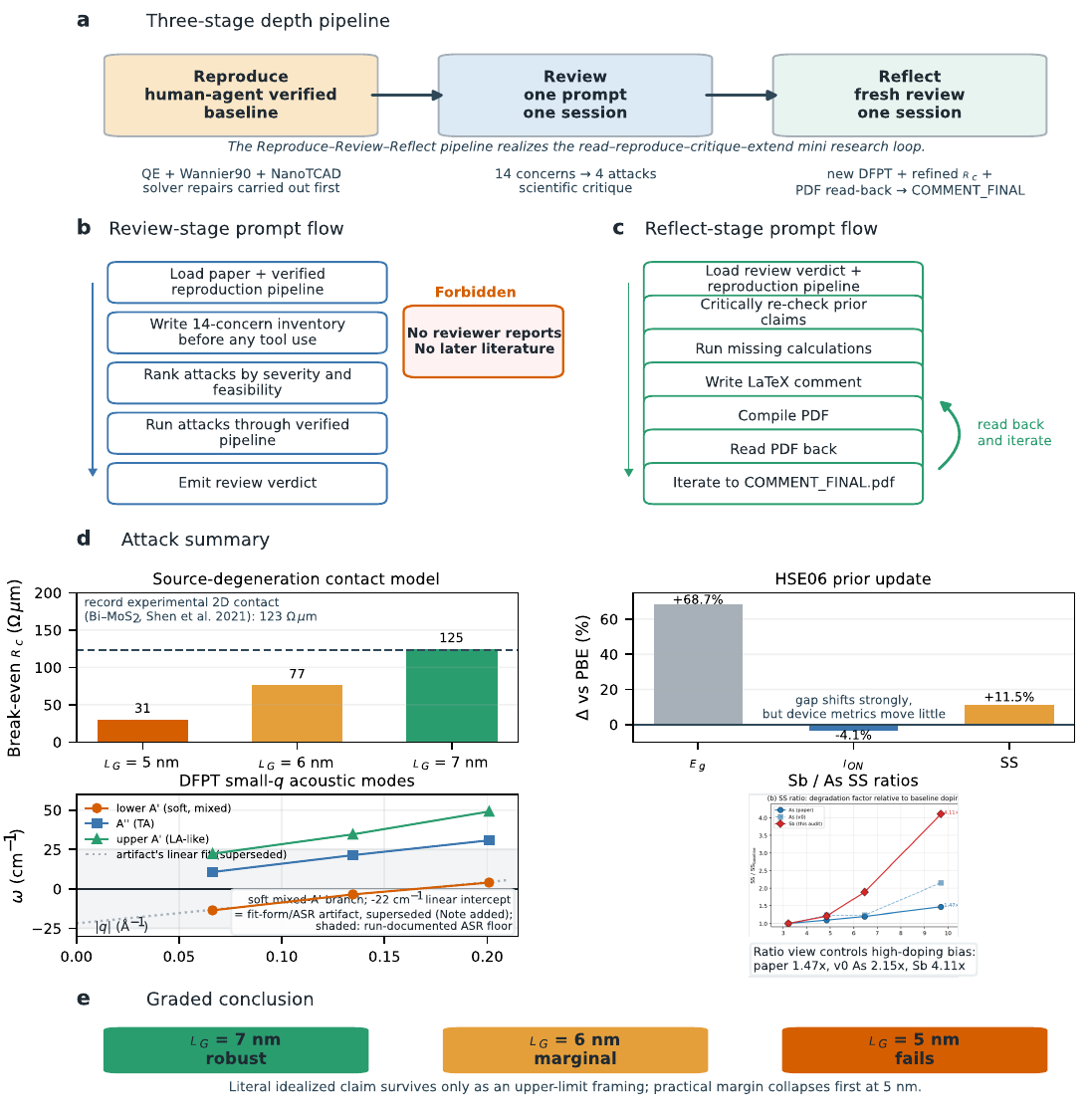}
\caption{\textbf{The Reproduce--Review--Reflect pipeline applied to Pizzi~2016.}
\textbf{(a)}~Three-stage flow: Reproduce (human--agent verified baseline across QE + Wannier90 + NanoTCAD) $\to$ Review (one session; 14-concern inventory, four attacks) $\to$ Reflect (fresh session; new DFPT, refined $R_c$, PDF read-back loop $\to$ COMMENT\_FINAL).
\textbf{(b)}~Review-stage prompt flow: load paper + verified pipeline $\to$ write 14-concern inventory before any tool use $\to$ rank attacks $\to$ run attacks $\to$ emit review verdict. Forbidden inputs: reviewer reports, subsequent literature.
\textbf{(c)}~Reflect-stage prompt flow: load verdict $\to$ re-check claims $\to$ run missing calculations $\to$ write \LaTeX{} $\to$ compile $\to$ read PDF back $\to$ iterate to COMMENT\_FINAL.pdf.
\textbf{(d)}~Attack summary: source-degeneration $R_c$ = 31/77/125~$\Omega{\cdot}\mu$m at $L_\text{G}$ = 5/6/7~nm vs.\ the record experimental 2D contact, $123~\Omega{\cdot}\mu$m for semimetal--MoS$_2$~\cite{shen2021}; HSE06 prior update ($E_g$ +68.7\%, $I_\text{ON}$~$-$4.1\%, SS +11.5\%); DFPT small-$q$ acoustic (soft mixed-A$'$ flexural-derived branch; the artifact's $-22$~cm$^{-1}$ linear intercept is an ASR/fit-form artifact, superseded per the Note added in \S\ref{sec:reflect}; panel redrawn from the run's raw frequencies); Sb/As SS ratios (paper 1.47$\times$, pipeline As 2.15$\times$, Sb 4.11$\times$; Sb-vs-As acceleration 1.91$\times$).
\textbf{(e)}~Graded conclusion: $L_\text{G} = 7$~nm robust, 6~nm marginal, 5~nm fails.}
\label{fig:fig3}
\end{figure*}

The depth pipeline (Fig.~\ref{fig:fig3}a) is a three-stage sequence on one paper: \textbf{Reproduce} once, build a verified end-to-end reproduction across all four codes; then \textbf{Review} and \textbf{Reflect}, two fresh autonomous agent sessions that, respectively, audit the paper against the verified pipeline and upgrade the audit into a scientific Comment.

\subsection{Reproduce: building a verified reference pipeline}

The Reproduce stage on Pizzi~2016 was carried out as human--agent collaboration, almost entirely because of the legacy state of the transport solver the paper relies on. NanoTCAD~ViDES~\cite{marian2023,fiori2005} was last updated in 2016 and carries silent failure modes that require instrumentation to surface (engineering-challenges summary in Appendix~\ref{app:engineering}). The human work was mainly tool repair; the downstream multi-code reproduction across QE $\to$ Wannier90 $\to$ instrumented NanoTCAD --- pseudopotential selection, convergence parameters, and calibration against the paper's published figures of merit --- was the agent's. This is a one-time tool-engineering cost, not a recurring per-paper cost; the resulting verified pipeline is what the autonomous Review and Reflect stages inherit.

\subsection{Review: a 14-concern inventory, four attacks, and comparison with human peer review}
\label{sec:review}

The Review agent received the paper plus the verified reproduction pipeline under a ``physics-first, tools-second'' prompt (Appendix~\ref{app:methods} M3, Appendix~\ref{app:prompts}): write a structured concerns inventory before any tool use, then pursue attacks. This ordering is itself the result of a prompt ablation --- a tools-first inventory, obtained by reading paper and tooling knowledge together as in scale mode, produces a narrower list of attacks anchored on what the tools easily compute (Appendix~\ref{app:methods} M3). The Review agent ran 2~h~11~min and issued 239 tool calls. It produced a \textbf{14-concern physics inventory} (up from 5 under the tools-first ablation), ranked by severity and feasibility, attempted \textbf{four computational attacks} end-to-end (contact resistance, HSE+SOC bandgap, an analytical flexural-phonon argument, a Sb doping sweep), and ran \textbf{nine further cross-checks} validating individual paper claims. We summarize two attacks --- one headline-challenging (contact resistance), one null-result control where the agent's own initial intuition is overturned by the calculation (HSE prior update).

\textbf{Attack~A --- contact resistance.} The paper sets source/drain contact resistance to zero in one sentence and never quantifies it. The agent post-processed the paper's own $f_T$ and $\tau$ with two analytical series-resistance models, finding break-even contact resistances at $L_\text{G} = 5$~nm below even the experimental record for a 2D contact, $123~\Omega{\cdot}\mu$m for semimetal--MoS$_2$~\cite{shen2021} --- an author-supplied benchmark; the agent's own figures (${\sim}100$--$1000~\Omega{\cdot}\mu$m, Bi--MoS$_2$ ${\sim}120$) came from parametric memory in the blind inventory phase, and the source-degeneration bound is benchmark-independent; the Reflect-stage refinement (\S\ref{sec:reflect}) later replaced these toy bounds with a source-degeneration model yielding \textbf{31/77/125~$\Omega{\cdot}\mu$m at $L_\text{G} = 5/6/7$~nm}. Under any realistic contact resistance, the $L_\text{G} = 5$~nm headline collapses.

\textbf{Attack~B --- HSE+SOC bandgap and the prior-update story.} Unprompted by paper or prompt, the agent proposed HSE~\cite{heyd2003,heyd2006} itself from the inventory, on the standard observation that PBE~\cite{perdew1996} underestimates gaps by 30--50\% and the na\"ive expectation --- it stated explicitly --- that a larger gap should give a steeper subthreshold slope through reduced band-to-band tunneling. It ran a full HSE06+SOC pipeline through Wannier and back into the device simulation. The gap rose by \textbf{+68.7\%}, as expected, but the device-level result inverted the prior: $I_\text{ON}$ shifted \textbf{$-$4.1\%}, local SS shifted \textbf{+11.5\%} (PBE $\to$ HSE), because the gate work-function offset absorbs approximately $\Delta E_g/2$ and the carrier-relevant conduction-band alignment is approximately conserved (Appendix~\ref{app:methods} M5). The agent recognised the inversion in its worklog and scoped the falsification to this attack rather than over-generalizing. \textbf{The agent originated the physics question, held an incorrect prior, and updated itself in writing once the calculation refuted it --- the failure mode that grounded execution is structurally protective against (\S\ref{sec:discussion}).}

The other two Review attacks bore on the headline indirectly. On flexural-phonon scattering, Review made an analytical argument that the paper's Takagi-formula treatment was inadequate but did not have time to back it computationally. On Sb doping, Review ran the sweep Pizzi~2016 omitted and obtained a qualitative scaling, but the final bracket did not fully enclose the $I_\text{OFF}$ target. Both open threads were picked up in Reflect (\S\ref{sec:reflect}).

\begin{figure*}[!tb]
\centering
\includegraphics[width=0.88\textwidth]{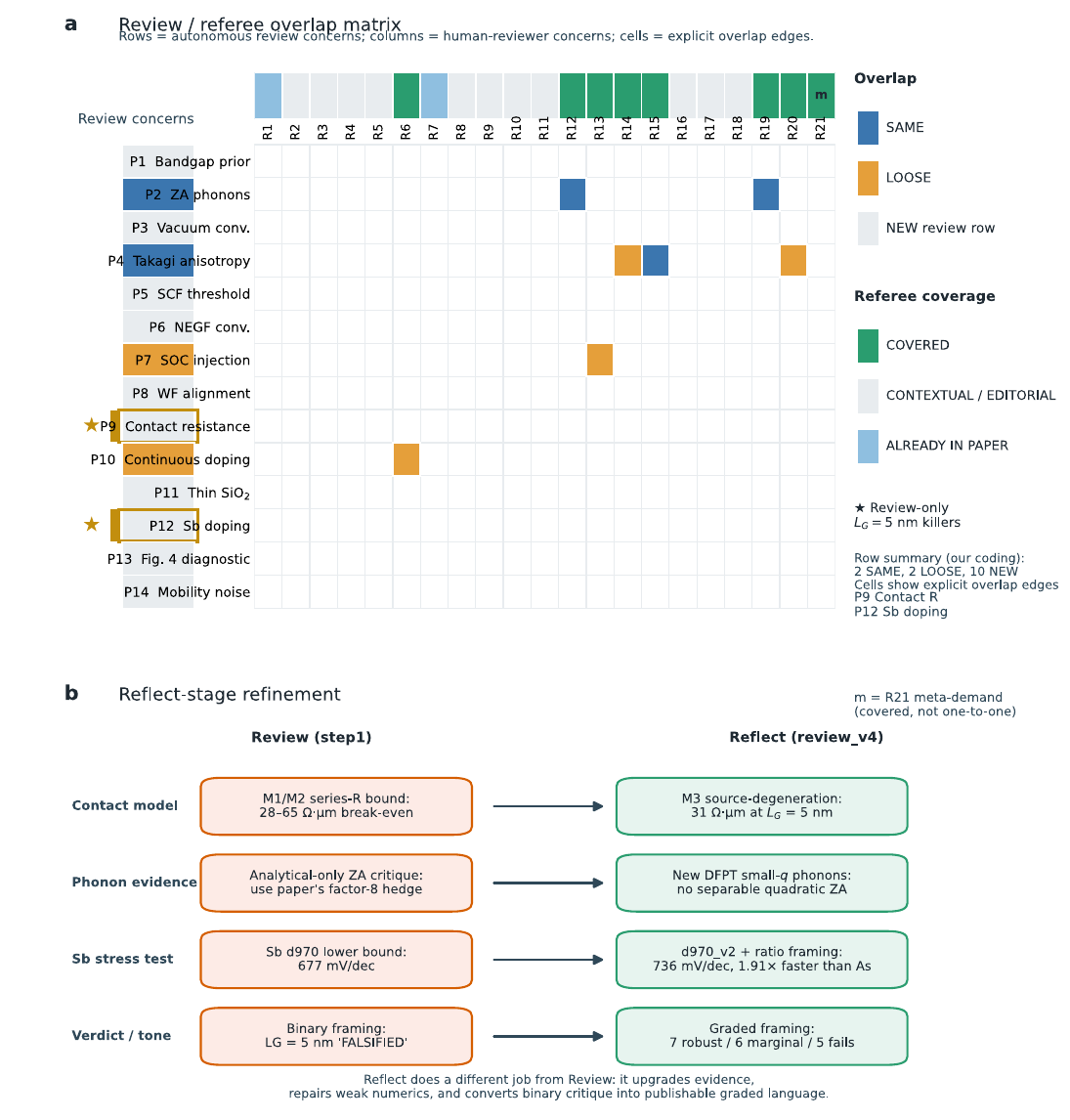}
\caption{\textbf{Review~$\leftrightarrow$~referee overlap and Reflect-stage refinement.}
\emph{Top:} sparse $14 \times 21$ overlap matrix, rows = Review concerns (P1--P14), columns = referee concerns (R1--R21). Row-level coding (Review side, SAME/LOOSE/NEW) summarises each Review concern's strongest overlap class. Under our coding scheme: \textbf{2 SAME, 2 LOOSE, 10 NEW} (Review side); 8 COVERED / 11 EDITORIAL-or-CONTEXTUAL / 2 ALREADY-IN-PAPER (referee side); Appendix~\ref{app:overlap} gives the full rule-set. Gold outline and star mark the two Review-only $L_\text{G} = 5$~nm headline-challenging attacks, P9 (contact resistance) and P12 (Sb doping).
\emph{Bottom:} four paired Review~$\to$~Reflect refinement rows: contact modelling M1/M2 bound (28--65~$\Omega{\cdot}\mu$m) $\to$ M3 source-degeneration (31~$\Omega{\cdot}\mu$m at $L_\text{G} = 5$~nm); analytical ZA-phonon critique $\to$ first-principles DFPT; Sb stress test (d970 lower bound $\to$ d970\_v2 ratio framing, 1.91$\times$ faster than As); binary ``FALSIFIED'' verdict $\to$ graded (7 robust / 6 marginal / 5 fails).}
\label{fig:fig4}
\end{figure*}

\textbf{Comparison with the published peer review.} \emph{Nat.\ Commun.}\ publishes its peer review files; Pizzi~2016 received 21 reviewer concerns across two rounds. The Review agent was forbidden from reading reviewer reports or subsequent literature, creating an orthogonal information asymmetry: the agent has computational reach without literature fluency; reviewers have the converse. Figure~\ref{fig:fig4} shows the overlap as a sparse $14 \times 21$ matrix. Under our coding scheme, two Review concerns map directly to referee concerns (SAME), two are loose family overlaps (LOOSE), and the remaining ten are new --- the exact counts depending on a small number of judgment calls (Appendix~\ref{app:overlap}). The more load-bearing observation is qualitative and robust: \textbf{both $L_\text{G} = 5$~nm headline-challenging attacks --- contact resistance (P9) and Sb doping (P12) --- are Review-only.} To our knowledge this is the first end-to-end existence proof that an autonomous agent, starting from a published paper, produces grounded findings on its central claim that human peer review did not. \textbf{Agent and human peer review have orthogonal attack surfaces, and their union is strictly larger than either alone}; humans do not run calculations during review.

\subsection{Reflect: from Review verdict to a grounded follow-up artifact}
\label{sec:reflect}

Review left two scientific threads open --- the analytical phonon argument and the incomplete Sb doping bracket --- alongside the usual marks of a single audit run: tone problems (``FALSIFIED everything''), an arithmetic error in the phonon mean-free-path scaling, and a binary verdict where a graded one is more honest. We handed the Review corpus to a fresh agent under a boilerplate prompt (Appendix~\ref{app:prompts}): review the audit, verify what is verifiable, run the calculations Review did not, produce a scientific Comment. The agent converged end-to-end on a publication-shaped artifact without human intervention, closing the audit by reading its own compiled PDF back into context and iterating (Fig.~\ref{fig:fig3}c; Appendix~\ref{app:reflect}).

The output, \textbf{COMMENT\_FINAL.pdf}, is a six-page scientific Comment (reproduced as-is in Appendix~\ref{app:comment}) with figures, references, methods, and an explicit list of open questions; substantive improvements are in Fig.~\ref{fig:fig3}d,e and summarised as $L_\text{G} = 7$~nm robust / 6~nm marginal / 5~nm fails. None of the three pieces of work Reflect added is interpolation from the paper's parameters. The \textbf{source-degeneration contact model} is a different physical model from the Review-stage toy bound: it identifies that the paper's own $f_T$ and $\tau$ already constrain the maximum extractable transconductance, derives the resulting break-even contact resistance from first principles, and gives a ceiling independent of any external 2D-contact-resistance benchmark. The \textbf{DFPT phonon calculation} replaces Review's analytical Takagi argument with first-principles DFPT on freestanding arsenene, resolving a soft, flexural-derived low-lying branch of mixed A$'$ character that the analytical treatment was structurally incapable of capturing. Buckled arsenene lacks the horizontal mirror $\sigma_h$, so the flexural branch cannot be symmetry-excluded from first-order transport scattering as in planar 2D crystals, and it lies far below the LA-like branch the paper's Takagi treatment assumes ($|\omega| \leq 14$ vs.\ 22--49~cm$^{-1}$ over the sampled $|q| = 0.066$--$0.201$~\AA$^{-1}$). The artifact's headline number, a $-22$~cm$^{-1}$ $q \to 0$ intercept, requires correction: it is a linear extrapolation of three raw \texttt{ph.x} frequencies without acoustic-sum-rule (ASR) enforcement. Rotational invariance makes a flexural-derived branch quadratic-dominated near $\Gamma$, so a linear fit mechanically yields a negative intercept, and the run's own output records a residual ASR violation (${\sim}5$--$25$~cm$^{-1}$) larger than the lowest measured frequency ($-13.6$~cm$^{-1}$). The intercept is a fit-form artifact, not evidence of dynamical instability (Note added below). The corrected reading strengthens the critique: the paper's LA-only treatment omits a soft mixed-character branch far below its assumed acoustic spectrum. The \textbf{Sb high-doping closing point} completes Review's sweep at the doping level that actually brackets the $I_\text{OFF}$ target, converting the qualitative scaling into a quantitative degradation ratio (Sb degrades 1.91$\times$ faster than As under matched doping). \textbf{The Comment is the form, but the content is grounded follow-up scientific work}: three classes of calculation that go beyond what Pizzi~2016 reported, packaged in the shape of an object that \emph{could} be submitted.

The four attacks occupy different evidential classes. The source-degeneration bound is derived from the paper's own $f_T$ and $\tau$, independent of any external benchmark or pipeline parameter; the HSE attack is a self-falsified control --- the agent's stated prior overturned by its own calculation; the DFPT phonon thread is first-principles but interpretation-sensitive, as the Note added below records; and the Sb ratio is a pipeline-dependent quantitative result with a matched-doping internal control (the matched As ladder).

\textbf{Note added (camera-ready).} During workshop review of this manuscript, a reviewer's reading-based assessment identified an interpretation error in the depth artifact --- the acoustic-branch characterization corrected above --- together with citation-fidelity lapses, corrected throughout this version. This is itself a datum for \S\ref{sec:discussion}: execution grounds numbers, but branch assignment, interpretation, and citation fidelity sit outside that protection, and here they were caught by human domain knowledge operating on the text. It is not evidence that reading-based LLM review could not catch them: this depth pipeline ran Review and Reflect as single sessions, without the session-breaking adversarial review loops that scaffolded systems can add. Whether such scaffolded review reliably catches this error class --- and how robustly --- is an open question we pose rather than answer. The corrected characterization strengthens the underlying critique of the original paper's phonon treatment, and the graded conclusion --- $L_\text{G} = 7$~nm robust, 6~nm marginal, 5~nm fails --- is unchanged: it rests on the contact-resistance and Sb-doping attacks.

\section{Discussion}
\label{sec:discussion}

\textbf{Computation as the central epistemic act.} The failure modes documented in the AI Scientist literature~\cite{lu2026,beel2025} --- hallucinated results, unreliable self-review, na\"ive idea generation --- are failures of the \textbf{blank-slate paradigm}, where the system generates content with no physical reference. Our results are for a different paradigm, \emph{grounded scrutiny}, where every step is anchored against an external physical reference: the paper's numbers (themselves validated against experiment or independent first-principles work) and the agent's own re-runnable simulations of the same physics. A hallucinated number does not survive execution: running the calculation either reproduces the target or does not, and the judge is the physics. Execution grounds numbers; it does not by itself protect setup choices, interpretation, or citation fidelity (\S\ref{sec:reflect}, Note added). \emph{Grounding in published physical science is structurally protective against hallucination modes any blank-slate generative system must contend with --- not as a feature of our implementation, but as a property of grounded scrutiny itself.} The 96.6\% execution requirement at scale and the HSE prior-update at depth are the same finding at two scales: physical reasoning is not the bottleneck; \emph{running the thing} is. The thesis, stated precisely, is descriptive and falsifiable: in corpora of published --- hence already human-reviewed --- computational-physics papers, the substantive methodological concerns that remain are overwhelmingly execution-bound (85/88 here). This is in part a selection effect, and it is also the deployment condition: reading-detectable errors are largely filtered by conventional review before publication, so an autonomous second pass adds value precisely where reading stops. We make no counterfactual claim that reading-based review --- human or LLM --- cannot surface critique; the Note added in \S\ref{sec:reflect} is a live counterexample on our own artifact. The framing is testable: on pre-review corpora such as preprints, the reading-detectable fraction should rise.

\textbf{What grounded scrutiny demands.} The capabilities exercised end-to-end here are exactly the three the introduction identifies as discriminators of real physical science: physics reasoning that cannot be reduced to interpolation (pseudopotential, functional, $k$-mesh, convergence criterion); multi-scale execution on decades-mature scientific software (QE $\to$ Wannier90 $\to$ NEGF, with the paper's prose and figures as the only spec); and verifiability against re-runnable physical ground truth at every step. Every step is physics-level reasoning, not text-level pattern matching.

\textbf{Harness, not model.} Every limitation we encountered traces to the harness, along four engineering-addressable axes. (i)~\textbf{Knowledge.} Two text files unlock cognitive scope the model already has (\S\ref{sec:scale}); required-reading envelopes are a primitive solution, and a richer structured tool layer is the natural next step~\cite{huang2026a}. (ii)~\textbf{Tools.} The Reproduce-stage cost on Pizzi~2016 was itself a tool-maturity problem; a tool layer with well-engineered solvers amortises that cost across future depth cases. (iii)~\textbf{Compute resource management and planning.} Agents over-subscribe cores, leak subprocesses on long sessions, and even when told to spend unlimited time bias toward narrowing scope and finishing --- both reflect using a short-task coding CLI as a long-horizon research orchestrator. (iv)~\textbf{Visual capability.} Agents miss topological differences between band-structure plots, extract values from plots with large error, and never look back at their own figures --- a multimodal-reading problem the harness can solve through programmatic figure checks and structured plot data. Of these, (i) is established by a controlled ablation (\S\ref{sec:aggregate}); (ii)--(iv) are diagnoses from the trace audit and the depth case, not ablations. Scaling grounded scrutiny means scaling the harness along all four axes, not waiting for a new model.

\textbf{Limitations.} Critique coding is single-coder and post hoc (released rule-set and worked examples enable independent recoding; Appendix~\ref{app:methods} M6, Appendix~\ref{app:phase}). Only harness axis (i) is backed by a controlled ablation. The depth result is one case on one deliberately selected paper --- an existence proof, no cross-paper generalization claimed. One model, one harness, one code ecosystem; corpus selection effects (published, open-access, QE-primary, 2010--2024). The soft 2--4~h budget bounds attempted scope; open-loop visual behaviour limits self-audit of figures.

\textbf{Outlook.} Grounded scrutiny is the natural first stage of a full real-world research loop --- read, reproduce, critique, extend. The depth case here already does the third and a measure of the fourth; an agent that originates its own scientific question and writes a full paper beyond an incremental Comment is the natural extension --- demonstrated in companion work~\cite{huang2026c}, which runs the full corpus-to-manuscript loop with anchor reproduction as its calibration mechanism; the two pipelines are complementary halves of one program. Separately, a direct practical use already exists: deployed alongside human peer review, pre-verified reproduction pipelines plus autonomous Review and Reflect stages give the literature a second epistemic mode it does not currently have --- not ``was this paper read carefully'' but ``was this paper \emph{run}'' --- and could function as a fully automated complement to the existing process.

\bibliographystyle{icml2026}
\bibliography{references}

\newpage
\appendix
\onecolumn

\section{Methods}
\label{app:methods}

\textbf{M1.\ Corpus construction.} The scale-mode corpus was assembled from an OpenAlex~\cite{priem2022} snapshot (March~2026) of all papers citing Quantum ESPRESSO, yielding \textbf{30{,}779 candidates}. We filtered in three passes. (i)~Asset availability: \textbf{10{,}926 entries} retaining both full-text PDF and supplementary material downloadable without institutional authentication. (ii)~Open-access verification: \textbf{4{,}708 entries} with a publicly licensed route, of which \textbf{2{,}641} were verified by direct download. (iii)~Primary-tool filter: an LLM classifier (Claude Sonnet~4.6) read each paper and identified \textbf{708 entries} in which Quantum ESPRESSO is the primary computational tool. From this pool we drew a stratified random sample plus earlier development batches used for prompt and knowledge-envelope iteration. Work proceeded in five batches: B1--B4 were development runs used to converge the prompt, the knowledge envelope, and the harness; B5 comprises two independent 50-paper runs on separate machines, of which the Mac run is the controlled single-machine production run quoted in \S\ref{sec:emergent}; the 12 papers common to both B5 machines form the cross-machine check. The 90-paper production set used in Appendix~\ref{app:vignettes} is the deduplicated union of unique OA papers across B3--B5 (B3: 17; B4: 15, a subset of B3's set; B5: 50 per machine; deduplication yields 90). Corpus-wide reproduction-quality and critique-phase statistics are computed on the full deduplicated 111-paper OA corpus; the ${\sim}42\%$ paper-level critique rate is 47/111. The controlled knowledge-ablation referenced in \S\ref{sec:scale} is the B3$\to$B4 comparison on a fixed 15-paper open-access subset (same model, same compute, only the knowledge files and the consult-before-refuse rule added). The cross-machine reproducibility figure uses a 12-paper subset run on a second workstation with no shared state, filesystem, or prompt history.

\textbf{M2.\ Harness, prompt, and required-reading envelope.} Both scale and depth regimes use the same harness: Claude Code CLI + Claude Opus~4.6, on a single workstation with ${\sim}12$ CPU cores and ${\sim}64$~GB RAM. The agent interacts with the system only through bash shell calls. \textbf{No central tool layer} is exposed: no MCP server, no library wrapper. This is a deliberate choice to make trace analysis tractable and to keep the harness honest about what the model can do with shell access alone.

Scale mode wraps this harness in a Python outer loop that iterates over the corpus, spinning up a fresh agent per paper with a 2--4~h wall-clock soft cap framed as a flexibility budget; scope reductions are declared in the verdict. Each agent receives five required-reading files that define the envelope: \texttt{HOUSE\_RULES.md} (resource-management discipline, time-budget discipline, honesty conventions); \texttt{READING\_GUIDE.md} (structure of the reading summary and the reproduction-target list); \texttt{VERDICT\_FORMAT.md} (JSON schema for the per-paper verdict); \texttt{INDEX.md} (Quantum ESPRESSO command idioms --- \texttt{pw.x}, \texttt{ph.x}, \texttt{epw.x}, \texttt{epsilon.x}, \texttt{open\_grid.x}, \texttt{pp.x}); and \texttt{PSEUDOPOTENTIALS.md} (pseudopotential-selection heuristics: SSSP vs PseudoDojo, NLCC compatibility with hybrid functionals, SOC-capable families). Core Wannier90 ecosystem documentation is also included; NanoTCAD~ViDES and other transport solvers are not. The converged production prompt emerged across B1--B4 and is stable for B5. The B3$\to$B4 ablation on a fixed 15-paper OA subset (same model, same compute, only the knowledge files and consult-before-refuse rule added) is the controlled ablation reported in the main text.

\textbf{M3.\ Depth-mode pipeline.} Depth mode uses the same CLI + Opus~4.6 harness but without a Python outer loop: three sequential agent sessions on one paper (Reproduce, Review, Reflect), each under its own prompt. The Review and Reflect prompts enforce a ``physics-first, tools-second'' structure: inventory of concerns written before any tool use, attack ranking by severity and feasibility, explicit forbidden-inputs clauses excluding reviewer reports and subsequent literature. The physics-first ordering was converged on after a tools-first ablation in which paper and tooling knowledge were read together before planning (the scale-mode ordering): under that ordering the concerns inventory anchored on what the available tools easily computed and stayed narrow, yielding 5 concerns against the 14 of the physics-first run on the same paper.

The Reproduce stage on Pizzi~2016 was carried out over roughly one week of intense human--agent debugging cycles. A summary of the tooling arc is in Appendix~\ref{app:engineering}: the verified QE + Wannier90 + instrumented NanoTCAD toolchain produced by this stage is what Review and Reflect inherit.

\textbf{M4.\ Grading, in-scope definition, and verdict schema.} Each paper is graded on a four-tier scale: \textbf{T4}\,=\,all in-scope claims the agent attempted reproduce within its self-declared tolerance; \textbf{T3}\,=\,all in-scope claims reproduce qualitatively with small numerical offsets and the underlying mechanism preserved; \textbf{T2}\,=\,some in-scope claims reproduce while others fail or are abandoned; \textbf{T1}\,=\,no in-scope claim reproduces. \emph{In-scope} is relative to the agent's own declared scope at the start of the session: what its QE installation, knowledge envelope, time budget, and available compute allow it to attempt. Claims the agent declares out-of-scope are excluded from the grade. The verdict JSON schema includes, for each claim, the paper-side value, the agent-side value, a numerical deviation, a scope classification, and a free-text reasoning field.

\textbf{M5.\ Gate work-function mechanism behind the HSE prior update.} The na\"ive ``larger gap $\to$ steeper SS'' expectation is borrowed from band-to-band-tunneling (TFET) device physics and does not transfer here. In the long-channel limit the subthreshold slope is set by gate electrostatics alone ($\mathrm{SS} = \ln 10 \cdot (kT/q)(1 + C_\mathrm{d}/C_\mathrm{ox})$, with no bandgap or effective-mass dependence); at $L_\text{G} = 5$~nm the additional degradation channel is source-to-drain tunneling through the gate barrier, whose transmission depends on barrier shape and band-edge effective mass --- not on the gap itself. Operationally, under a functional change PBE $\to$ HSE the gate work-function offset required to hold the same $I_\text{OFF}$ target shifts by approximately $\Delta E_g/2$, the carrier-relevant OFF-state conduction-band alignment is approximately conserved, and the observed ($I_\text{ON}$ $-$4.1\%, SS +11.5\%) shift is attributed in the run's worklog to the transverse band-edge effective mass and OFF-state charge density dominating over the gap-driven WKB exponent change; the device remains source-to-drain-tunneling-limited under both functionals ($\mathrm{SS} \gg 60$~mV/dec).

\textbf{M6.\ Phase A/B/C/D classification.} Each of the 88 substantive critiques was classified post hoc by a deterministic, re-runnable keyword classifier over the session's terminal verdict artifact, with Phase-A assignment additionally requiring either corroborating concern language in the pre-execution reading summary or the absence of any executed calculation; the two original Phase-A rows and a 10-paper sample were then hand-verified against reading summaries and worklogs. The classification is single-coder-audited and post hoc; the released rule-set, classifier, and four worked examples (one per phase, Appendix~\ref{app:phase}) enable independent recoding, and the boundary rule there is conservative against overstating reading-only detection. A camera-ready re-audit corrected the single clearest boundary error --- a concern tagged post-execution whose session never executed a calculation --- via the rule's own no-calculation clause, giving corrected counts A = 3, B = 22, C = 49, D = 14 (85/88 = 96.6\%). An independent blinded LLM recoding of all 88 rows (one fresh judge per row, given only the rule-set, the row's reading summary, its verdict artifact, and the mechanical fact of whether any calculation executed) agreed with the corrected boundary on 66/88 rows ($\kappa = 0.16$), with every disagreement in one direction: the blinded judges coded as reading-detected those rows where the reading summary already stated the concern's substance and execution, in their reading, only quantified it --- cases the released boundary rule codes as execution-bound. The execution-bound share is therefore coding-sensitive: 96.6\% under the released (corrected) instrument, 71.6\% under the blinded recoding; the bolded headline uses the corrected instrument, and we report the recoding as the measured sensitivity of this single-coder boundary. Throughout, ``execution-dependent'' is a positional statement --- the concern first surfaces after calculation execution in the session record --- not a causal claim about what produced the critique. The finer B/C/D sub-splits carry more keyword noise than the A-vs-non-A boundary and should be read as indicative.

\textbf{M7.\ Data availability.} All production inputs, outputs, traces, required-reading envelopes, prompts, verdict JSONs, the verified Pizzi~2016 reproduction pipeline, COMMENT\_FINAL.pdf, the cross-machine reproducibility subset, development-batch artifacts (marked as excluded from quantitative results), figure-generation scripts, and the Pizzi~2016 Reproduce-stage debug journal will be released upon publication under CC-BY~4.0; the Comment itself is reproduced as-is in Appendix~\ref{app:comment}.

\section{Workflow diversity gallery}
\label{app:gallery}

\begin{figure}[h]
\centering
\includegraphics[width=0.95\textwidth]{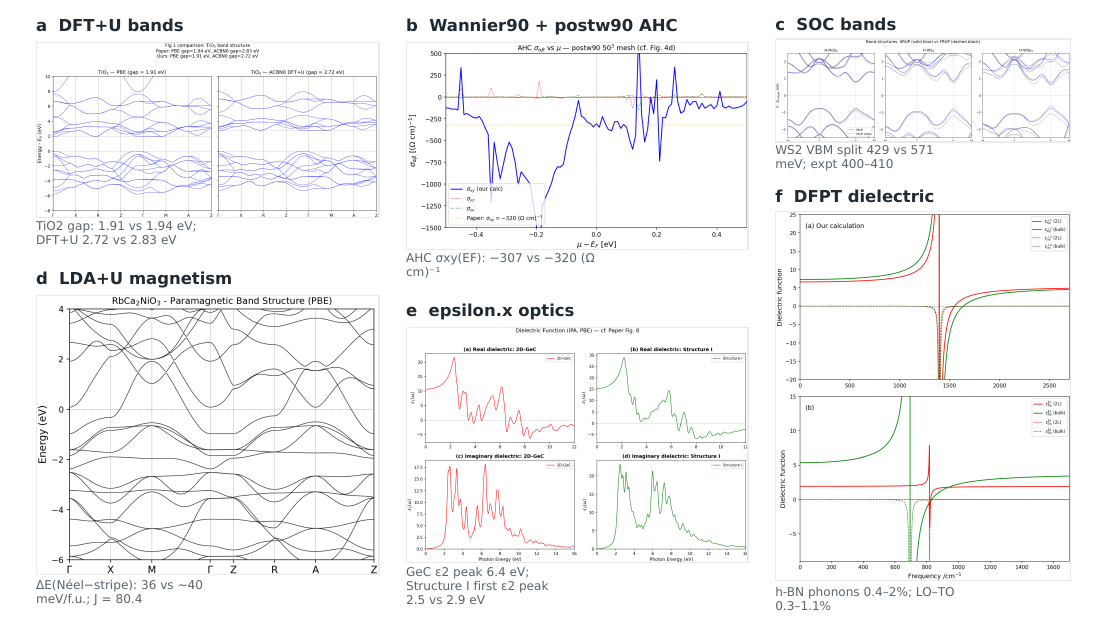}
\caption{\textbf{Workflow diversity gallery.} Six autonomous agent-vs-paper comparisons in a three-column mosaic:
\textbf{(a)}~DFT+U bands, TiO$_2$ gap 1.91 vs 1.94~eV; DFT+U 2.72 vs 2.83~eV~\cite{agapito2015}.
\textbf{(b)}~Wannier90 + postw90 AHC, $\sigma_{xy}(E_F) = -307$ vs $-320$~$(\Omega\;\text{cm})^{-1}$~\cite{park2022}.
\textbf{(c)}~SOC bands, WS$_2$ VBM split 429 vs 571~meV; experiment 400--410~meV~\cite{reyesretana2016}.
\textbf{(d)}~LDA+U magnetism, $\Delta E(\text{N\'{e}el}-\text{stripe}) = 36$ vs ${\sim}40$~meV/f.u.; $J = 80.4$ vs 80--90~meV~\cite{nomura2020}.
\textbf{(e)}~\texttt{epsilon.x} optics, GeC $\varepsilon_2$ peak 6.4~eV; Structure~I first $\varepsilon_2$ peak 2.5 vs 2.9~eV~\cite{islam2021}.
\textbf{(f)}~DFPT dielectric, h-BN phonons 0.4--2\%, LO--TO 0.3--1.1\% agreement~\cite{wang2016hbn}.
Each panel was produced in a single agent session with no human intervention.}
\label{fig:figA1}
\end{figure}

Figure~\ref{fig:figA1} gives six autonomous agent-vs-paper comparisons, each produced in a single agent session with no human intervention, spanning DFT+U, Wannier90+postw90 anomalous Hall, spin--orbit coupling, LDA+U correlated magnetism, \texttt{epsilon.x} optics, and DFPT dielectrics.

\section{Phase A/B/C/D classification rule-set}
\label{app:phase}

Each substantive critique was classified against the first stage at which it became a paper-side concern in the trace.

\begin{itemize}
  \item \textbf{Phase A:} the concern is present during reading, before execution beyond paper/guide inspection.
  \item \textbf{Phase B:} the concern surfaces while diagnosing a failed or misbehaving calculation.
  \item \textbf{Phase C:} the concern emerges after a successful calculation is compared to the paper.
  \item \textbf{Phase D:} the concern becomes substantive only after a second independent calculation or cross-check changes the confidence state.
\end{itemize}

The corrected phase totals in the audited substantive ledger are $A=3$, $B=22$, $C=49$, $D=14$, for 88 substantive critiques (one row reclassified at camera-ready per the rule-set's no-calculation clause; Appendix~\ref{app:methods} M6). Of those, 85/88 (96.6\%) are execution-dependent. The reading-only ceiling is 2/111 = 1.8\%: of the three Phase-A rows, one is a corrupted-source artifact rather than a genuine catch, one is a genuine physics catch, and one is a genuine reproducibility-omission catch.

\begin{table}[h]
\centering
\small
\caption{Worked examples of the phase-coding rule-set.}
\label{tab:phase-examples}
\begin{tabularx}{\textwidth}{l l l X}
\toprule
Phase & PID & Short label & Why it belongs there \\
\midrule
A & QEL-2020-01088 & Phosgene +142~Ry & Adsorption-energy number is already physically impossible at reading time; no calculation is needed to know that 142~Ry is not a plausible molecular adsorption energy. \\
B & QEL-2012-00155 & Hybrid-NLCC failure mode & The critique emerges while trying to make the HSE workflow behave: PP/NLCC incompatibility and resulting gap collapse are discovered in failure diagnosis. \\
C & QEL-2016-00265 & WS$_2$ SOC splitting & The concern appears only after a successful fully relativistic calculation yields 429~meV against the paper's 571~meV. \\
D & QEL-2019-00174 & Fe$_2$TaAl phonons & The stronger paper-side concern only appears after cross-checking the $\Gamma$-point optical frequencies across two PP families, both of which refute the paper's quoted scale. \\
\bottomrule
\end{tabularx}
\end{table}

\textbf{Boundary rule.} Boundary cases were resolved by the \emph{first decisive surfacing} rule. If a reading-stage intuition later became a genuine critique only after execution, the phase is execution-bound, not reading-bound. Likewise, if an initial comparison-stage discrepancy later received a stronger independent confirmation, the critique remains Phase~C unless the second computation is what made it credible in the first place. This conservative coding biases against overstating reading-only detection.

\section{Review~$\leftrightarrow$~referee overlap coding rule}
\label{app:overlap}

The overlap coding uses the following rule on each Review-side concern (rows P1--P14) against each human referee concern (columns R1--R21).

\begin{itemize}
  \item \textbf{SAME:} same physical objection family, same load-bearing mechanism, and close enough in scope that a referee could reasonably regard them as the same criticism.
  \item \textbf{LOOSE:} same broad family or device-nonideality direction, but materially different mechanism or narrower/wider scope.
  \item \textbf{NEW:} no meaningful overlap on the review side.
\end{itemize}

On the referee axis, \textbf{COVERED} means directly addressed by a SAME or LOOSE Review concern, \textbf{EDITORIAL/CONTEXTUAL} means outside the Review session's load-bearing physics-audit frame, and \textbf{ALREADY-IN-PAPER} means a reviewer-raised issue that had already been incorporated into the published version later read by Review. Under this coding scheme the row summary is \textbf{2 SAME, 2 LOOSE, 10 NEW}, and the referee-side summary is \textbf{8 COVERED, 11 EDITORIAL/CONTEXTUAL, 2 ALREADY-IN-PAPER}. The two SAME mappings are P2$\leftrightarrow$R12/R19 (ZA-phonon / ballistic-regime vulnerability) and P4$\leftrightarrow$R15 (Takagi oversimplification). P7$\leftrightarrow$R13 is deliberately only LOOSE: the referee asked whether SOC was present at all, whereas Review questioned the transport-pipeline injection. P9 (contact resistance) and P12 (Sb doping sensitivity) remain the two Review-only $L_\text{G} = 5$~nm headline-challenging attacks. R21 (a reviewer sensitivity-analysis meta-demand across the whole inventory) is kept as a covered \emph{meta} column rather than a one-to-one cell.

\textbf{Review-side inventory P1--P14.} P1 PBE gap / effective-mass bias; P2 ZA-phonon exclusion undermining ballistic justification; P3 20~\AA{} vacuum for charged 2D supercell; P4 Takagi isotropic/parabolic mismatch with anisotropic valleys; P5 SCF-threshold misreport; P6 NEGF $k$-mesh/energy-step convergence not demonstrated; P7 SOC injection into ViDES; P8 work-function-shifted $I_\text{OFF}$ normalisation; \textbf{P9 contact resistance set to zero}; P10 continuous-sheet doping approximation; P11 sub-1~nm physical SiO$_2$; \textbf{P12 doping sensitivity of the gate barrier}; P13 Fig.~4 ``no tunneling charges'' diagnostic conflation; P14 ${\sim}5\%$ mobility-tensor numerical noise.

\textbf{Referee-side concerns R1--R21.} Opening texts (abridged): R1 tight-binding model construction; R2 phosphorene/As/Sb detail; R3 puckered As allotrope; R4 indirect-gap discussion; R5 why large mobilities; R6 realistic disorder sources; R7 finite-$T$ effects; R8 As/Sb look too similar; R9 Table~I effective-mass interpretation; R10 why stop at $L_G=7$; R11 Table~II explanation; R12 ZA flexural phonons in buckled crystals; R13 SOC in DFT/TB pipeline; R14 electron--phonon anisotropy; R15 Takagi oversimplified; R16 Wannier-TB related-work crediting; R17 experimental realization; R18 shorten long paragraphs; R19 unresolved ZA concern (round 2); R20 mobility method remains unconvincing (round 2); R21 sensitivity-analysis meta-demand (round 2).

\section{Reflect-stage iteration summary}
\label{app:reflect}

The depth pipeline separates into a long Review session and a shorter Reflect session. Session metrics: Review (v15) = 239 tool calls, 131.4~min; Reflect (review\_v4) = 234 tool calls, 53.7~min. The reflect-stage document loop itself used four compile invocations and two explicit PDF read-backs before \texttt{COMMENT\_FINAL.pdf} was frozen. The iteration history is four drafts: (1)~Review verdict (14-concern inventory, four attacks, nine verified claims); (2)~Reflect v1/v2 recalibration; (3)~Reflect v3 locks in graded conclusion and the DFPT replacement plan; (4)~Reflect v4 adds the source-degeneration contact model, the arsenene DFPT calculation, the clean d970\_v2 Sb point, and the final six-page Comment with PDF read-back.

\begin{table}[h]
\centering
\footnotesize
\caption{Key depth-stage numerical anchors referenced in the main text.}
\label{tab:depth-anchors}
\begin{tabularx}{\textwidth}{X X X}
\toprule
Quantity / attack & Review stage & Reflect stage (final) \\
\midrule
HSE06+SOC gap $E_g$ (arsenene, $L_G=5$~nm) & 1.4681 eV (PBE+SOC) $\to$ 2.4767 eV & same; $+68.7\%$ relative shift \\
$I_\text{ON}$ PBE $\to$ HSE propagation & 2903~A/m $\to$ 2783~A/m ($-4.1\%$) & confirmed \\
Local SS PBE $\to$ HSE & $+11.5\%$ & confirmed \\
Contact resistance model M1 / M2 / M3 at $L_G=5$~nm & 27.7 / 65.2 / --- $\Omega{\cdot}\mu$m & M3 source degeneration: 31~$\Omega{\cdot}\mu$m \\
$R_c$ break-even at $L_G=5/6/7$~nm & --- & 30.95 / 76.75 / 125.30 $\Omega{\cdot}\mu$m \\
Sb degradation (d970 / baseline, verified pipeline) & 677~mV/dec lower bound & 736 / 179, ratio 4.11 \\
As degradation (d970 / baseline, verified pipeline) & --- & 272 / 126, ratio 2.15 \\
Sb-vs-As SS acceleration & qualitative & $4.11 / 2.15 = 1.91\times$ \\
DFPT arsenene soft mixed-A$'$ branch & analytical only & $\omega = -13.6\ldots{+4.0}$~cm$^{-1}$ over $|q| = 0.066$--$0.201$~\AA$^{-1}$; the $-21.89$~cm$^{-1}$ linear intercept is superseded as an ASR/fit-form artifact (\S\ref{sec:reflect}, Note added) \\
\bottomrule
\end{tabularx}
\end{table}

The HSE result is the depth-side null-result control: it validates a severe upstream bandgap correction while showing that the specific device-level headline remains comparatively robust to that correction once the work-function alignment is re-calibrated. The source-degeneration $R_c$ lands close to Review's harsher M1 bound, not its more forgiving M2 bound. The DFPT result resolves a soft long-wavelength mixed-A$'$ branch far below the LA-only Takagi picture the original paper uses; the apparent linearity and negative $q \to 0$ intercept in the artifact are artifacts of a three-point linear extrapolation without ASR enforcement (\S\ref{sec:reflect}, Note added), while the branch's softness and mixed character stand. The Sb ratio converts the qualitative Review-stage scaling into a quantitative degradation ratio.

\section{Per-phase representative vignettes}
\label{app:vignettes}

We include one vignette per phase. PIDs are our internal identifiers.

\textbf{Phase~A: phosgene adsorption energy (+142~Ry), QEL-2020-01088.} The reading summary listed adsorption energies of 142.09~Ry for phosgene on a [5,0] CNT and 136.96~Ry on a BN nanotube. That is orders of magnitude beyond plausible molecular adsorption; typical adsorption energies are on the order of $10^{-3}$--$10^{-1}$~Ry. This is the rare genuine reading-only physics catch in the corpus.

\textbf{Phase~B: hybrid-NLCC failure mode in GaAs/InAs defects, QEL-2012-00155.} The paper's main headline depends on hybrid-functional bulk gaps and downstream defect energetics. In reproduction, the critique surfaced while debugging the HSE setup itself: PseudoDojo PPs with NLCC produced a near-zero hybrid gap, while SG15 no-NLCC PPs restored the expected qualitative behaviour. The final verdict confirms the paper's qualitative need for hybrids (PBE gives a zero GaAs gap, as the paper states) while showing that the narrow-gap side of the setup is extremely PP-sensitive. Textbook Phase-B: the critique is born inside failure diagnosis.

\textbf{Phase~C: WS$_2$ SOC splitting, QEL-2016-00265.} The fully relativistic reproduction gave a WS$_2$ valence-band splitting of 429~meV against the paper's 571~meV, a 24.9\% discrepancy. The key feature is not that the calculation failed --- it succeeded, reproduced the qualitative physics, and then disagreed with the paper in a scientifically informative way. The worklog attributes the discrepancy to the paper's custom RRKJ ultrasoft fully-relativistic pseudopotentials overestimating W-based SOC splittings; the reproduced value is closer to experiment (400--410~meV). Phase~C is the modal discovery point in the corpus.

\textbf{Phase~D: Fe$_2$TaAl/Fe$_2$TaGa phonons, QEL-2019-00174.} The initial QE DFPT result already suggested a large discrepancy in the optical-mode scale, but the stronger critique only became credible after a second PP-family cross-check. GBRV ultrasoft and PSLibrary PAW calculations both gave $\Gamma$-point optical frequencies in the ${\sim}180$--$330$~cm$^{-1}$ range, whereas the paper quotes ${\sim}510$--$680$~cm$^{-1}$ for Fe$_2$TaAl and ${\sim}470$--$591$~cm$^{-1}$ for Fe$_2$TaGa. The paper's qualitative dynamical-stability claim survives, but the quantitative optical-mode scale looks systematically wrong by roughly a factor of two. The second independent computation is what converts a suspicious mismatch into a substantive critique.

\textbf{Critique taxonomy.} The OA-v2 audit tracks nine categories, of which eight substantive ones compose the 88-row ledger: \texttt{paper\_error} (21), \texttt{paper\_bias} (20), \texttt{agent\_outperforms} (14), \texttt{missing\_correction} (11), \texttt{method\_limitation} (9), \texttt{pp\_artifact} (5), \texttt{paper\_omission} (5), \texttt{experimental\_benchmark} (3). The ninth, \texttt{convergence\_issue} (60 instances), is tracked operationally but excluded from the 88-row substantive ledger, because a convergence failure is not by itself a paper-side methodological concern.

\textbf{Denominator conventions.} Paper-level substantive-critique rates across denominators: full deduplicated OA corpus 47/111 (42.3\%); deduplicated production papers only 47/90 (52.2\%); clean single-machine B5 Mac run 30/50 (60.0\%). The value 88 is not a paper-level rate; it is the critique-instance count used for phase and category analyses.

\section{Trace-level behavioural observations}
\label{app:traces}

The trace-behaviour audit used the 61 session-trace files of the B5 production machine: 50 completed papers, one partial session, and 10 rate-limit-aborted stubs that contain no tool calls. Three compact findings anchor the main-text claims.

\textbf{Load-then-execute is the aggregate signature.} Across the first 20 tool calls of each of the 51 live sessions (1020 calls; the 10 stubs contribute none), 586/1020 calls (57.5\%) were direct file reads and 664/1020 (65.1\%) were read-like actions once closely related lookup actions were included. No session begins by launching QE. Reading summaries are written in 50 of the 51 live sessions, and 51/61 traces reference one --- the single live session without a summary file discusses the summary in-session but never writes it. Among sessions with scorable late behaviour, 49/50 show zero paper rereads after execution starts. Comparison-figure compliance in the audited B5 figure total is 107/107. The safe claim is procedural: once the summary became mandatory, production traces show a stable read~$\to$~plan~$\to$~execute discipline with durable on-disk planning state. The ``load-then-execute'' framing is a tool-call statement, not a cognition statement: after the initial load, agents may still be attending to earlier paper text internally even when no explicit reread appears in the trace.

\textbf{Filesystem as external memory under compaction.} Compaction affected 10/61 sessions (16.4\%), with 11 total compaction events. Zero sessions re-read the paper after the last compaction; only 2 sessions re-opened the reading summary. The emergent strategy is not ``reconstruct the paper state from scratch'' but ``continue from on-disk artifacts'' --- generated inputs, outputs, worklogs, and the reading summary already written to disk. We did not prompt for this behaviour.

\textbf{Open-loop visual behaviour.} In 0/61 sessions does an agent generate a figure, open it, notice a problem, and regenerate. Agents reliably produce \emph{requested} visual artifacts but do not yet treat their own visual output as a self-auditing feedback channel. This is the basis for treating visual verification as a harness limitation rather than a reasoning limitation.

\section{Prompt and knowledge-envelope structural overview}
\label{app:prompts}

The production prompts and the required-reading envelope are structured artifacts rather than monolithic instructions. We give a one-paragraph structural description of each rather than verbatim text; complete files will be released upon publication under CC-BY~4.0.

\textbf{Scale prompt (B5, converged production form).} A short orchestration layer (${\sim}30$ lines) that references the five required-reading files by name, sets the four-step workflow (load paper $\to$ write reading summary $\to$ execute calculations with worklog $\to$ emit verdict), defines a single output directory per paper, and states the flexibility time budget. It does not ask for critique; it asks for reproduction artifacts, comparison figures, and a structured verdict.

\textbf{\texttt{HOUSE\_RULES.md} (${\sim}30$ lines).} Resource-management and process discipline: one-calculation-at-a-time, core-count discipline, full-paper-load requirement, comparison-figure-deliverable requirement, honesty conventions (when-in-doubt-declare-out-of-scope rather than silently drop a claim), and time-awareness scaffolding.

\textbf{\texttt{READING\_GUIDE.md} (${\sim}40$ lines).} Specifies the structured pre-compute planning output: section-by-section summary, a reproduction-target list with per-target scope/feasibility annotations, and explicit scope declarations for items the agent will not attempt. Forces a complete paper understanding before execution.

\textbf{\texttt{VERDICT\_FORMAT.md} (${\sim}50$ lines).} JSON schema for the per-paper verdict, including paper-side value, agent-side value, deviation, scope classification, and free-text reasoning for each attempted claim. Standardizes outputs for post hoc aggregation.

\textbf{\texttt{INDEX.md} and \texttt{qe\_executables.md} (${\sim}600$ lines combined).} Verified QE and Wannier90 capability map: executable idioms (\texttt{pw.x}, \texttt{ph.x}, \texttt{epw.x}, \texttt{epsilon.x}, \texttt{open\_grid.x}, \texttt{pp.x}), known gotchas for \texttt{pw2wannier90.x} k-point format, \texttt{postw90.x} Fermi-energy handling, adaptive Berry meshes, and \texttt{open\_grid.x} use in hybrid+SOC workflows. Encodes working idioms beyond the small subset agents spontaneously recall.

\textbf{\texttt{PSEUDOPOTENTIALS.md} (${\sim}200$ lines).} PP decision tree covering XC-matching, library choice (SSSP, PseudoDojo, SG15, GBRV, PSLibrary), NLCC compatibility with hybrid functionals, SOC-capable families, and workflow-specific PP constraints.

\textbf{Depth-mode Review and Reflect prompts (${\sim}150$ and ${\sim}120$ lines).} Both enforce a physics-first, tools-second structure: inventory of concerns written before any tool use, attack ranking by severity and feasibility, explicit forbidden-inputs clauses excluding reviewer reports and subsequent literature. The Reflect prompt additionally requires the compile-PDF-read-back-and-iterate loop and specifies the Comment-shaped output.

\textbf{B3$\to$B4 knowledge ablation.} On the same 15 OA papers, same model family, same 12-core compute budget, adding \texttt{INDEX.md} + \texttt{PSEUDOPOTENTIALS.md} + the consult-before-refuse rule. Aggregate: attempted quantitative claims 44~$\to$~47 (+7\%); within-5\% claims 40/44 (91\%)~$\to$~44/47 (94\%); mean absolute deviation 2.2\%~$\to$~1.3\% ($-41\%$); phonon workflows attempted on 3/15~$\to$~5/15 papers (20\%~$\to$~33\%); false ``QE cannot do this'' refusals eliminated on paired cases. The point is mechanism, not a large-$N$ performance claim: the agent did not become more intelligent; it stopped declining workflows already available in its local toolchain.

\section{Engineering challenges summary for the Reproduce stage}
\label{app:engineering}

The Reproduce stage on Pizzi~2016 was dominated by tool maturity differences across the QE / Wannier90 / NanoTCAD~ViDES stack. We summarize the recurrent challenge classes without reproducing the day-by-day journal; the full day-by-day notes will be released upon publication under CC-BY~4.0.

\begin{enumerate}
  \item \textbf{Legacy-runtime port.} The 2016-era transport-solver wrappers expected a legacy Python~2 interpreter and project-local conventions that did not survive unchanged into the 2026 environment. A dedicated runtime, wrapper repair, and an explicit rebuild path were prerequisites for any device result to be trustable.
  \item \textbf{Spinless-TB versus SOC-spinor factor-of-two mismatch.} The SOC spinor Wannier Hamiltonian traces both spin channels, while legacy transport code paths assumed a spinless tight-binding Hamiltonian and applied an extra factor of two. The correction had to be propagated consistently through the Landauer current path and the charge feed to the Poisson loop; the latter was the more consequential of the two because it distorted every SCF iteration. Several of these bugs had numerically compensating effects on the paper's published geometries, making disentanglement a systematic alternate-geometry exercise rather than a direct diagnostic.
  \item \textbf{Hard-coded lead-context assumptions.} The lead self-energy code implicitly assumed a minimum contact-pad headroom. In the principal-layer wrapper the effective lead context occupied four super-slices; at some short-pad settings this consumed essentially the full pad length, leaving no bulk-like lead region and distorting the device behaviour. The assumption was not visible from the Python interface; it was surfaced by reading the C source and then checked numerically by a pad-size diagnostic.
  \item \textbf{Silent-failure traps in the legacy core.} Memory-management issues on long $k$-point sweeps and a 600-iteration SCF ceiling could return numerically untrustable results without a clean exception. Source-level patches and a tracked-SCF wrapper converted silent solver behaviour into logged diagnostics.
  \item \textbf{Transport-truncation ambiguity.} The paper's ``58 neighbours'' language left genuine ambiguity about how much long-range Wannier hopping the device model retained, which triggered a principal-layer detour and a bit-identity unit-test suite verifying equivalence to the atomic reference implementation.
\end{enumerate}

\textbf{Methodological lessons.} Two lessons survive into the final paper story. First, multiple bugs were simultaneously present and partly cancelled on the publication geometry; the pipeline therefore passed through a deceptive intermediate regime where some observables looked nearly correct for the wrong reasons. Alternate geometries and continuation paths were not redundant checks; they were necessary to distinguish genuine agreement from accidental cancellation. Second, ``read everything from scratch'' repeatedly acted as a recovery mechanism: several decisive breakthroughs came from forcing a fresh read of the code path or the paper after the prior explanatory frame had become too sticky. The verified QE + Wannier90 + instrumented NanoTCAD toolchain produced at the end of this stage is what Review and Reflect inherit.

\textbf{Provenance of COMMENT\_FINAL.pdf.} The six-page Comment produced autonomously by Reflect has the following provenance chain, audited end-to-end: Review-stage verdict and worklog $\to$ Reflect-stage new calculations (source-degeneration contact model, arsenene DFPT, d970\_v2 Sb point) $\to$ \LaTeX{} draft $\to$ four compile invocations $\to$ two explicit PDF read-back events $\to$ final six-page PDF. No human editing was applied to the PDF body itself.

\medskip
\noindent\textbf{Release note.} Complete experimental artifacts, including full prompts, required-reading envelope files, per-paper traces, the NanoTCAD~ViDES debug journal, and the complete 9-category critique catalog, are available upon publication under CC-BY~4.0.

\section{The Reflect-stage Comment, reproduced as-is}
\label{app:comment}

The document below is the artifact produced autonomously by the Reflect session, reproduced without modification; its provenance chain is audited in Appendix~\ref{app:engineering}. It is included as primary data for the claim that a single unsupervised session can carry an audit through to a complete, submission-form scientific document --- the composition, figures, typesetting, and PDF iteration are the agent's own, for the reader to judge. Its scientific content is not warranted: it contains the acoustic-branch mischaracterization corrected in \S\ref{sec:reflect} (Note added) and may contain further issues; where this paper's main text and the artifact differ, the main text supersedes.

\includepdf[pages=-]{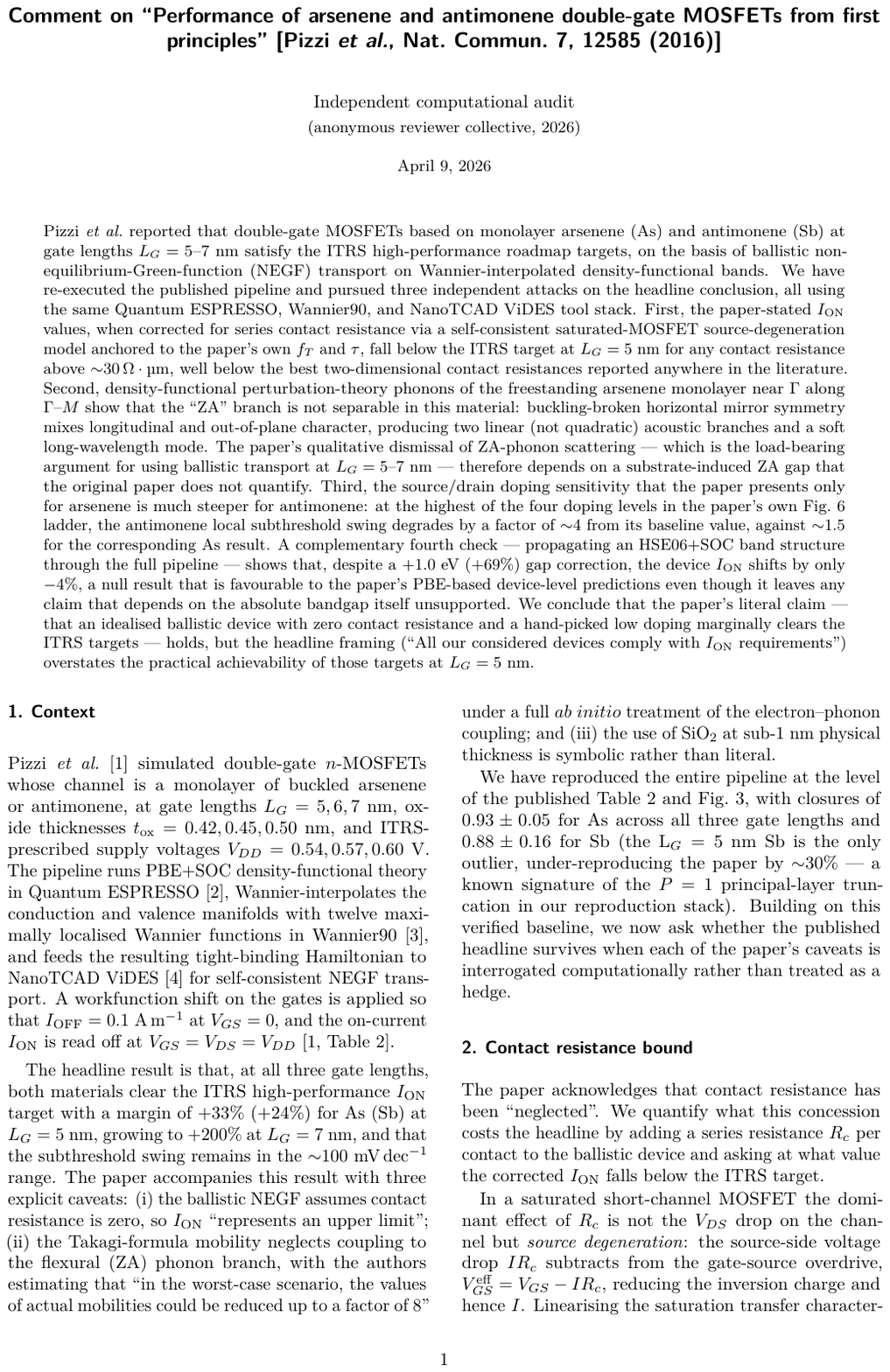}

\end{document}